\documentclass[pdftex,twocolumn,epjc3,preprintnumbers,amsmath,amssymb]{svjour3}          
\RequirePackage[T1]{fontenc}

\smartqed  

\RequirePackage{graphicx}
\RequirePackage{mathptmx}   
\RequirePackage{flushend}
\RequirePackage[numbers,sort&compress]{natbib}
\RequirePackage[colorlinks,citecolor=blue,urlcolor=blue,linkcolor=blue]{hyperref}

\journalname{Eur. Phys. J. C}
\usepackage{amsmath}
\usepackage{amssymb}
\usepackage[normalem]{ulem}
\usepackage{bm}
\usepackage{color}
\usepackage{xcolor}
\usepackage{soul}
\usepackage{graphicx}
\usepackage{xurl}
\usepackage[colorlinks,citecolor=blue,urlcolor=blue,linkcolor=blue]{hyperref}
\usepackage{multirow}
\usepackage{subfigure}
\usepackage{booktabs}
\usepackage{mathrsfs}
\usepackage{balance}
\setulcolor{yellow}

\begin{document}

\title{Constraints on Dark Matter Annihilation from Starlight-Induced Inverse Compton Emission in Omega Centauri}

\author{Guan-Sen Wang\thanksref{addr1,addr2}, Xu Pan\thanksref{addr3}, Lei Feng\thanksref{addr1,addr2,e2} 
}

\thankstext{e2}{e-mail: fenglei@pmo.ac.cn (corresponding author)}

\institute{Key Laboratory of Dark Matter and Space Astronomy, Purple Mountain Observatory, Chinese Academy of Sciences, Nanjing 210023, China\label{addr1} 
\and 
School of Astronomy and Space Science, University of Science and Technology of China, Hefei, Anhui 230026, China\label{addr2} 
\and    
School of Physics and Electrical-Electronic Engineering,  Aba Teachers  College, Aba, 623002,  China\label{addr3} 
}
\date{Received: date / Accepted: date}

\maketitle

\begin{abstract}
The search for the particle identity of dark matter (DM) remains one of the foremost challenges in modern physics. In this work, we derive new constraints on the DM annihilation cross section $\langle\sigma v\rangle$ by considering the inverse Compton (IC) scattering of DM annihilation injected $e^+e^-$ off the starlight radiation field in Omega Centauri. As a likely remnant core of a dwarf galaxy, Omega Centauri provides a dense stellar environment that enhances the predict IC signal. Using multi-wavelength observations from Chandra and Fermi-LAT, we establish upper limits on $\langle\sigma v\rangle$ over a wide DM mass range from MeV to TeV. Notably, in the tens-of-MeV range, our results surpass existing constraints by approximately two orders of magnitude, exceeding $\langle\sigma v\rangle \sim 10^{-31}~{\rm cm^3\,s^{-1}}$. At several GeV to the TeV DM masses range, our constraints also remain stronger than existing limits.  Our results demonstrate that starlight induced IC emission is an important component for indirect DM searches in dense stellar systems, establishing Omega Centauri as a powerful probe of the nature of DM.
\end{abstract}

\section{Introduction} \label{sec:introduction}

The existence of dark matter (DM) has been firmly established by a wide range of astrophysical and cosmological observations~\cite{Cirelli:2024ssz,1939LicOB..19...41B,Jiao:2023aci,Ou:2023adg,Labini:2023fmy,Clowe:2006eq,vanWaerbeke:2000rm,Massey:2007wb,Harvey:2015hha,Robertson:2016qef,Lewis:2006fu,Hanson:2009kr,Planck:2018lbu}. However, the identification of DM remains one of the most compelling challenges in modern physics. A well motivated candidate is Weakly Interacting Massive Particles (WIMPs)~\cite{Steigman:1984ac,Arcadi:2017kky,Lee:1977ua}, which can annihilate into Standard Model particles, including photons, neutrinos, and charged cosmic rays. This annihilation process offering a promising target for indirect detection. Traditionally, high energy gamma rays have been regarded as a primary target~\cite{Ibarra:2016fco,CermenoGavilan:2021ybc,Kumar:2016cum}, because they propagate freely and retain spatial information. However, this process is often model-dependent. In contrast, the $e^+e^-$ pairs produced by DM annihilation can interact with the surrounding astrophysical environment to generate secondary radiation across a broad electromagnetic spectrum~\cite{Buch:2015iya,Cirelli:2009vg,Kachelriess:2019oqu}. Over the past decades, various methods have been developed to explore the DM parameter space, such as searching for synchrotron emission in magnetized environments and inverse Compton (IC) scattering off the Cosmic Microwave Background (CMB) photons. While most researches have traditionally focused on the GeV-to-TeV mass range~\cite{Abdughani:2021pdc,Cui:2016ppb,Giesen:2015ufa,Cholis:2019ejx,Cuoco:2016eej,Nguyen:2024kwy,Baring:2015sza,Fermi-LAT:2016uux,MAGIC:2016xys,Fermi-LAT:2013thd,Muru:2025vpz}, the absence of conclusive signals has shifted attention toward the sub-GeV region~\cite{Knapen:2017xzo,Darme:2017glc,Boehm:2002yz,Boehm:2003hm,Boddy:2015efa,Gonzalez-Morales:2017jkx,Caputo:2022dkz,Yu:2025bwu}. In this lower mass regime, secondary signals in the X-ray bands have emerged as vital probes~\cite{Carenza:2022som,Boddy:2015fsa,Knodlseder:2016pey,Engel:2022bgx}. Some studies also adopt Synchrotron self Compton emission~\cite{Wang:2026khh}, or consider reacceleration effect of cosmic-ray to constrain the DM parameter space~\cite{Boudaud:2016mos,Zu:2021odn,Wang:2025jhy,Su:2024hrp}.

In previous works, the CMB is adopted as the primary or even sole target photon field of IC process~\cite{Porter:2008ve,Wang:2023sxr,Chen:2021rea}, but it often yields conservative limits. In dense stellar environments, the local starlight radiation field can exceed the CMB energy density by several orders of magnitude. Furthermore, the characteristic frequency of starlight is approximately three orders of magnitude higher than that of the CMB, shifting the resulting secondary IC emission into the X-ray and gamma-ray regimes where astrophysical backgrounds are more manageable. Thus, secondary radiation is expected to be a primary tool for uncovering the properties of DM.

In this work, we investigate the starlight induced IC process within a dense stellar environment to constrain DM. We select Omega Centauri, the most massive globular cluster in the Milky Way, as our research target. Its potential origin as the remnant nucleus of a stripped dwarf galaxy makes it a unique candidate for hosting a substantial DM component. Also, its high stellar density makes it an ideal target to test starlight-induced IC scattering.
We develop a theoretical framework that accounts for the realistic spatial and spectral distribution of starlight within Omega Centauri.
By analysing the multi-wavelength observations from the Chandra X-ray Observatory~\cite{Haggard:2009ts} and the Fermi Large Area Telescope (Fermi-LAT)~\cite{Brown:2019whs}, we constrain the cross section of DM annihilation across a wide DM mass range from MeV to TeV. Furthermore, we systematically examine the influence of astrophysical uncertainties, including the DM density profile and the spatial diffusion coefficient, on the derived constraints.

The remainder of this paper is organized as follows. In Sec.~\ref{sec:Omega_cen}, we introduce the properties of $\omega$ Cen, including its DM distribution and stellar radiation field. Sec.~\ref{sec:IC} details the theoretical modeling of the IC emission. The statistical methodology final results of constraining DM annihilation cross section are presented in Sec.~\ref{sec:constraints}. Finally, we provide a discussion and summarize our conclusions in Sec.~\ref{sec:conclusion}.

\section{Omega Centauri} \label{sec:Omega_cen}
Omega Centauri is the most massive globular cluster in the Milky Way. It has a stellar mass of $M \sim 10^{6}\,M_{\odot}$, a half-light radius of $r_{\rm h} \sim 7\,\rm pc$, and lies at a distance of $d \sim 5.4\,\rm kpc$ from the Earth~\cite{Harris:1996kt,Wang:2021hfb,Evans:2021bsh}. In this section, we describe the DM distribution of the cluster and construct the stellar radiation field.

\subsection{Dark Matter Distribution}
\label{sec:DMdistribution}
The exact amount of DM in Omega Centauri still remains uncertain~\cite{Garani:2023esk,Wang:2021hfb,Evans:2021bsh,Reynoso-Cordova:2022ojo}. Previous studies have reached different conclusions regarding the presence of an extended non-luminous mass component. While some analyses favor such a component~\cite{Wang:2021hfb,Evans:2021bsh}, yielding inferred $J$-factors of approximately $10^{22}$ to $10^{24}\,{\rm GeV^2\,cm^{-5}}$, others find no statistically significant evidence for a DM component and instead derive upper limits on its contribution~\cite{Reynoso-Cordova:2022ojo}. In addition, the non-luminous mass inferred from stellar dynamics may not all be due to DM, as stellar remnants could also contribute to it~\cite{Evans:2021bsh}. Despite these uncertainties, we adopt the Navarro–Frenk–White (NFW) profile~\cite{Navarro:1996gj} as a benchmark DM density distribution to investigate the potential signals from DM annihilation in Omega Centauri. The NFW profile is given by
\begin{equation}
\rho(r) = \frac{\rho_{\rm s} r_{\rm s}}
{r \left(1 + r/r_{\rm s}\right)^2},
\label{eqnfw}
\end{equation}
where $\rho_{\rm s}$ denotes the characteristic density and $r_{\rm s}$ is the scale radius of the DM halo.

To account for the uncertainty of actual DM distribution, we adopt three representative sets of NFW parameters taken from Ref.~\cite{Wang:2021hfb}, which are summarized in Tab.~\ref{tab:profile}. 
The corresponding $J$-factors for the three profiles are approximately $1.6\times10^{23}$, $5\times10^{23}$, and $1\times10^{23}\,\mathrm{GeV^2\,cm^{-5}}$, respectively. These profiles are used to illustrate the impact of the DM distribution on our results, rather than to represent the full range of possible DM distributions in Omega Centauri. The actual DM density profile may be either more or less concentrated than those considered here.

\begin{table}[!htb]
\renewcommand\arraystretch{1.5}
\centering
\begin{tabular}{c|c|c|c}
\hline\hline
 & Profile 1 & Profile 2 & Profile 3  \\
\hline
$r_{\rm s}$~(pc) & 1.63 & 1.00 & 2.00 \\
\hline
$\rho_{\rm s}~(M_{\odot}~\rm {pc^{-3}})$ & 7650.59 & 27860.50 & 4391.65 \\
\hline\hline
\end{tabular}
\caption{Parameters of the DM density profiles~\cite{Wang:2021hfb}.}
\label{tab:profile}
\end{table}

\subsection{Starlight Radiation Field}
\label{sec:starlight}

\begin{figure}
\centering 
\subfigure[]{\includegraphics[width=1\linewidth]{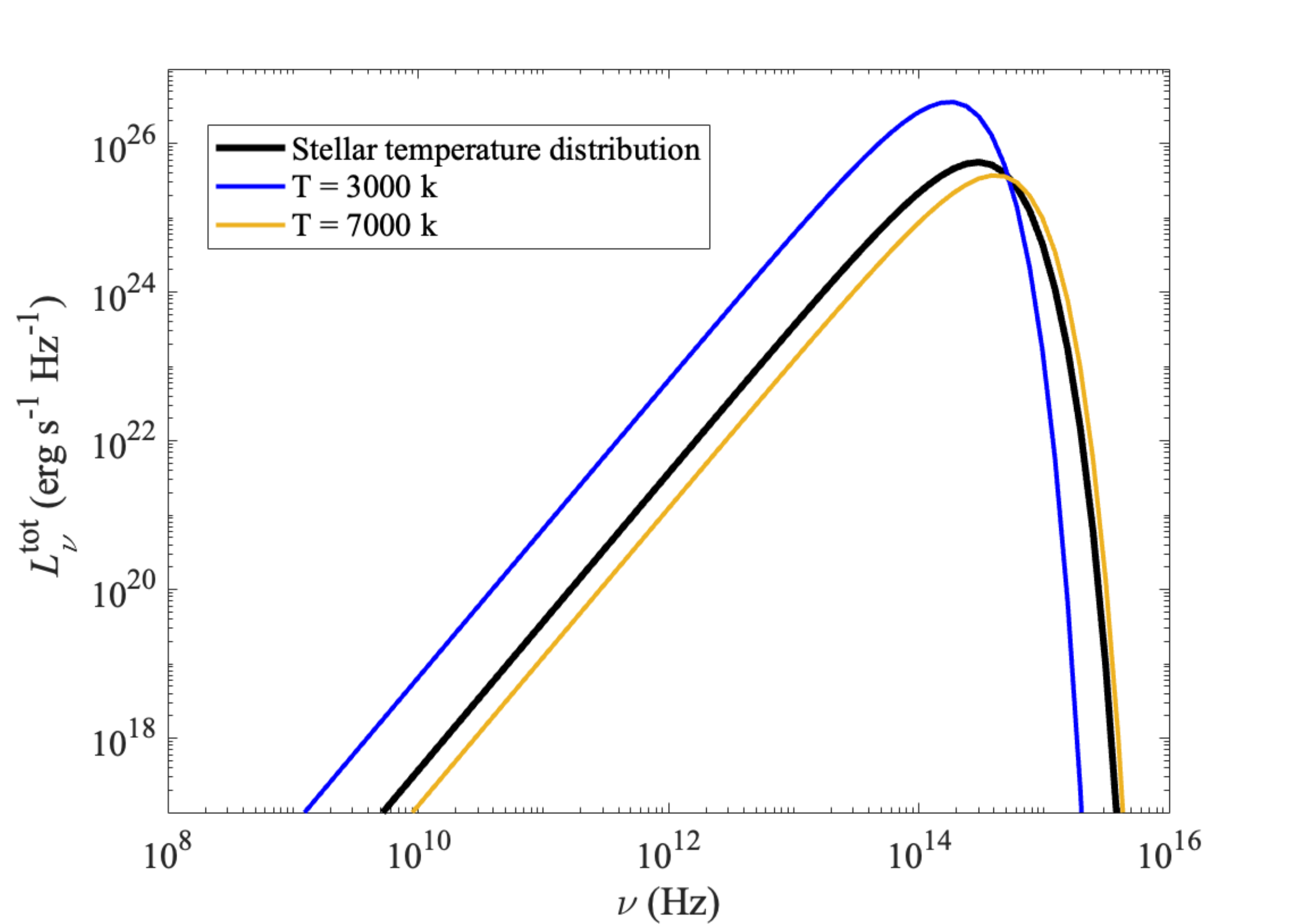}\label{fig:starlight_spectrum_a}} \\
\subfigure[]{\includegraphics[width=1\linewidth]{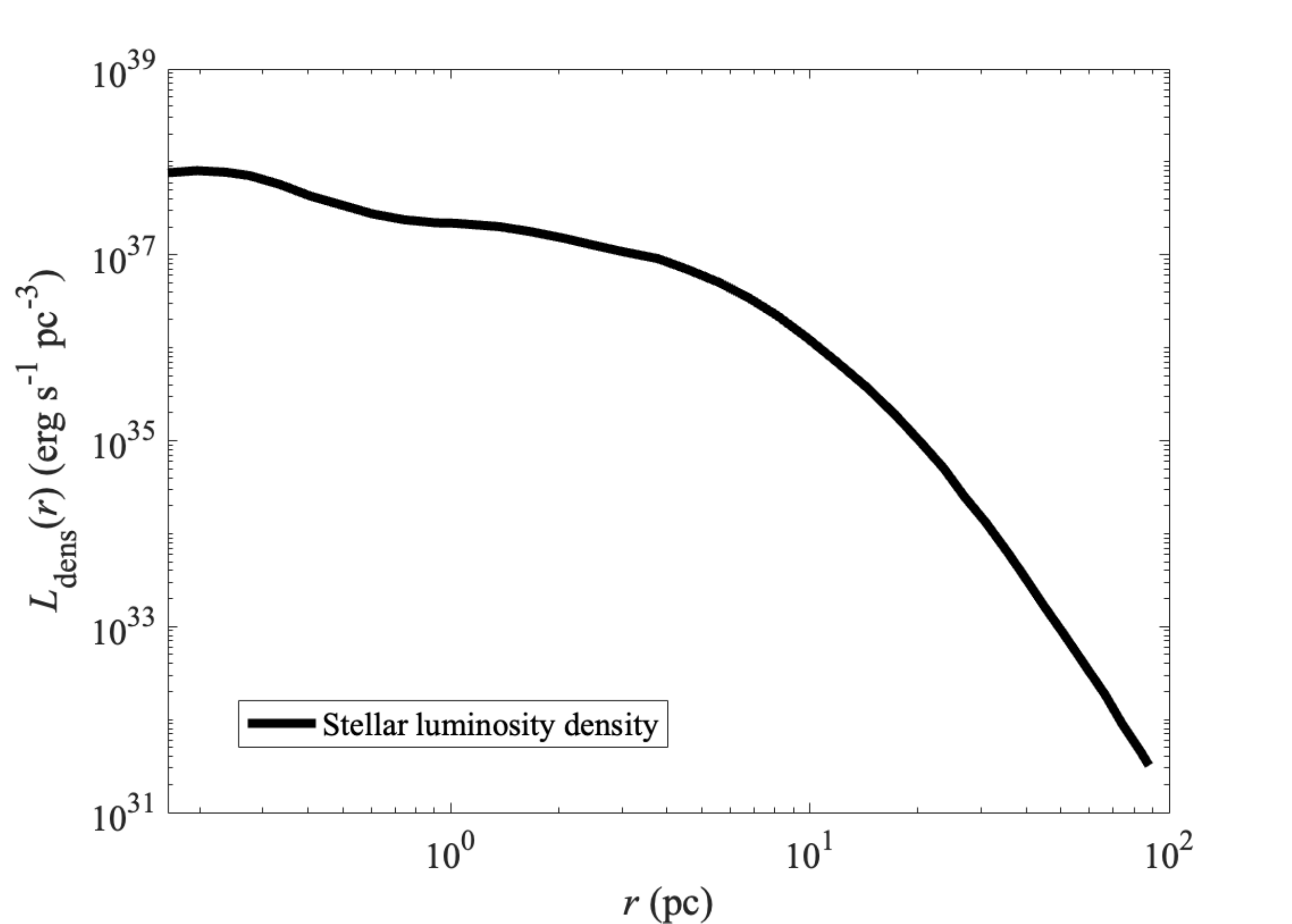}\label{fig:starlight_spectrum_b}} \\
\subfigure[]{\includegraphics[width=1\linewidth]{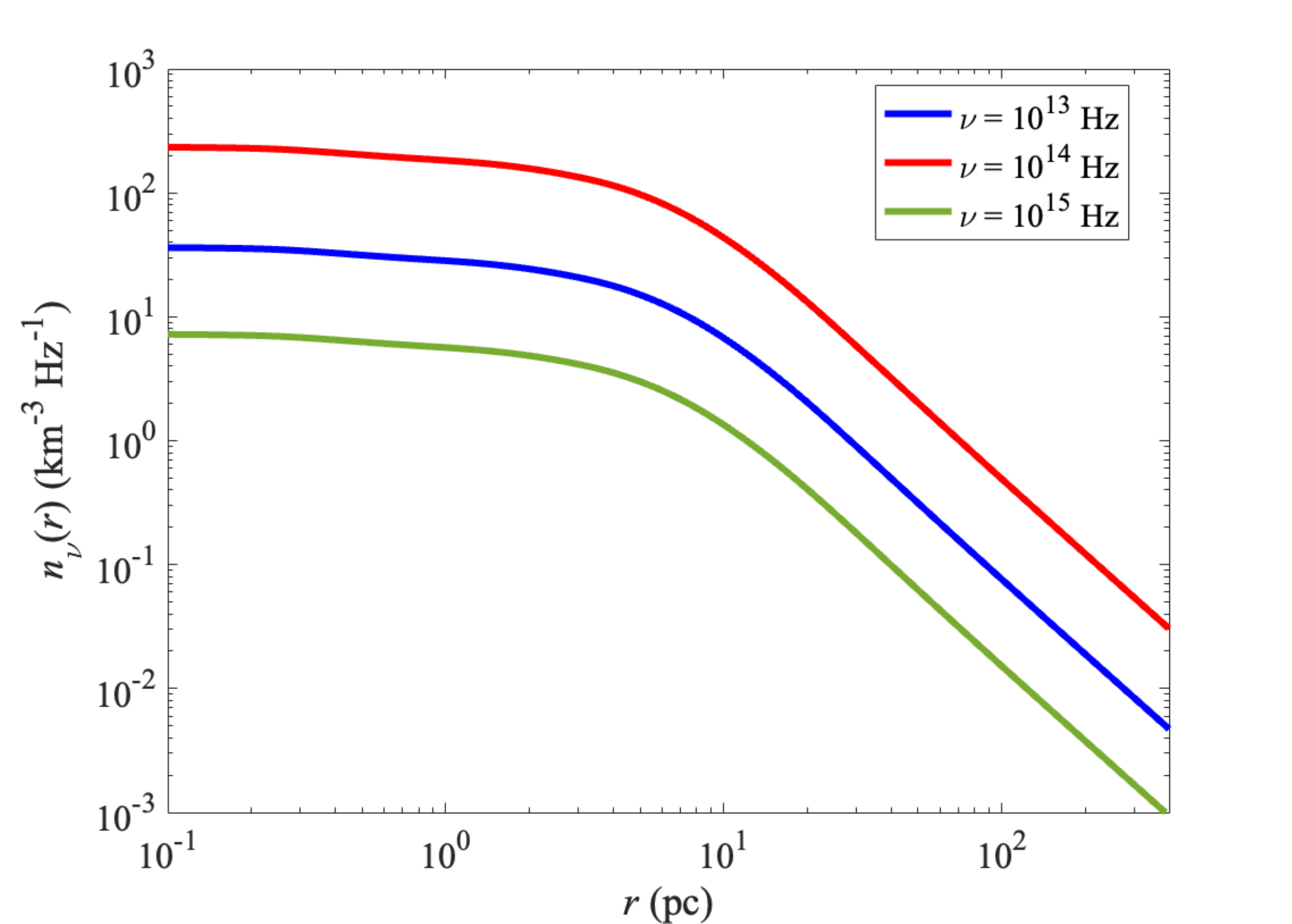}\label{fig:starlight_spectrum_c}} 
\caption{The starlight radiation field of Omega Centauri, with a stellar population age of $12\,{\rm Gyr}$~\cite{clontz2024omegacat} and metallicity $[{\rm Fe/H}]=-1.7$~\cite{Stanford:2006za}. The spectrum is normalized to the observed total $V$-band luminosity $L_V=1.086\times10^{6}\,L_\odot$. (a): The total starlight luminosity spectrum $L_\nu^{\rm tot}$ of the cluster (black line), with the spectra for the two cases in which all stars have effective temperatures of $3000\,{\rm K}$ (blue line) and $7000\,{\rm K}$ (yellow line) also shown for comparison. (b): The stellar luminosity density $L_{\rm dens}(r)$. (c): The photon number density $n_\nu(r)$ with three representative frequencies, $\nu=10^{13}$, $10^{14}$, and $10^{15}\,{\rm Hz}$, shown as the blue, red, and green curves, respectively.}
\label{fig:starlight_spectrum}
\end{figure}

In this subsection, we derive the starlight distribution of Omega Centauri by combining the observed luminosity data and stellar population synthesis.
The absolute magnitude of Omega Centauri, $M_V=-10.26$~\cite{SEDS_OmegaCen}, refers specifically to the $V$ band and corresponds to a $V$-band luminosity of approximately $L_V\simeq 1.086\times10^{6}\,L_\odot$. This luminosity is used to normalize the total stellar radiation field of the cluster. The $V$-band luminosity should be determined by integrating the stellar spectrum weighted by the $V$-band response function. For simplicity, we approximate this band by integrating over the frequency interval $4.5\times10^{14}\,\mathrm{Hz}\leq\nu\leq6.5\times10^{14}\,\mathrm{Hz}$, which is broader than the effective frequency range of the actual $V$-band response. This will overestimate the $V$-band luminosity and thus provides a conservative estimate of the stellar radiation field strength. For simplicity, we use the $V$ band as the sole normalization reference rather than fitting the spectrum to photometric measurements across multiple bands.
The spectral shape of the starlight radiation field is obtained using the PARSEC stellar population synthesis model~\cite{Bressan:2012zf}. We generate an isochrone table corresponding to a stellar population age of $12\,{\rm Gyr}$~\cite{clontz2024omegacat} and metallicity $[{\rm Fe/H}] = -1.7$~\cite{Stanford:2006za}. The isochrone tables provide the cumulative initial mass function and stellar effective temperatures. Assuming each star radiates as a blackbody, the total luminosity spectrum of the cluster $L_\nu^{\rm tot}$ can be obtained by summing the contributions from all stellar components.
The spectrum is normalized to the observed $V$-band luminosity, $L_V\simeq1.086\times10^6\,L_\odot$, and is shown in Fig.~\ref{fig:starlight_spectrum_a}.
We also consider two extreme cases in which all stars have the same effective temperature of either $3000\,{\rm K}$ or $7000\,{\rm K}$. Since cooler stars are less luminous and hotter stars are more luminous, we expect the actual cluster spectrum to be closer to the $7000\,{\rm K}$ case. We note that in the $3000\,{\rm K}$ case, the spectral peak lies below the $V$-band frequency range, resulting in a much higher total luminosity when normalized to the $V$-band luminosity. When wavelengths outside the \(V\)-band are taken into account, the \(3000\,\mathrm{K}\) spectrum is expected to differ significantly from the observations. Thus, this situation should be considered as an extreme illustrative case just for comparison.

Next, to describe the spatial distribution of the starlight radiation field, we assume that the spatial distribution of stellar luminosity follows the stellar mass distribution in Ref.~\cite{DSouza:2012grs}. The bolometric stellar luminosity density is then given by
\begin{equation}
L_{\rm dens}(r)=L_{\rm tot}\frac{\rho_\star(r)}{M_\star},
\end{equation}
where $L_{\rm tot}$ is the total luminosity of Omega Centauri, $\rho_\star(r)$ is the stellar mass density and $M_\star$ is the total stellar mass. The luminosity density is normalized such that its volume integral gives the total cluster luminosity, i.e.,
\begin{equation}
\int L_{\rm dens}(r)\,dV=L_{\rm tot}.
\end{equation}
The resulting spatial luminosity distribution is shown in Fig.~\ref{fig:starlight_spectrum_b}.
Assuming that the spectral shape of the starlight is independent of position, the photon production rate per unit volume and frequency is then given by
\begin{equation}
Q_\nu(r)=\frac{L_\nu(r)}{h\nu},
\end{equation}
where
\begin{equation}
L_\nu(r)=L_\nu^{\rm tot}\frac{\rho_\star(r)}{M_\star}
\end{equation}
is the luminosity density per unit frequency.

Here we assume the photon distribution of Omega Centauri is spherically symmetric. Then, the photon number density of the starlight radiation field is obtained by integrating the photon production rate over the cluster volume
\begin{equation}
n_\nu(r)=\frac{1}{c}\int \frac{Q_\nu(r')}{4\pi |r-r'|^2} dV'.
\end{equation}
The resulting photon number density distributions at three representative frequencies, $\nu=10^{13}$, $10^{14}$, and $10^{15}\,{\rm Hz}$, are shown in Fig. \ref{fig:starlight_spectrum_c}.

\section{Inverse Compton Emission from Dark Matter Annihilation}
\label{sec:IC}

In this section, we detail the theoretical framework for calculating DM-induced IC emission in Omega Centauri. First, we consider DM annihilation into relativistic $e^+e^-$ pairs. The injection rate of $e^+e^-$ at position $r$ and energy $E$ is given by
\begin{align}
Q(E,r)=\frac{\langle\sigma v\rangle\rho_{\chi}^2(r)}{2m_{\chi}^2}\frac{{\rm{d}}N}{{\rm{d}}E},
\label{esource_IC}
\end{align}
where $m_\chi$ is the mass of DM particles, $\langle\sigma v\rangle$ is the thermally averaged annihilation cross section, and $\rho_{\chi}(r)$ is the DM density profile as shown in Eq.\,\eqref{eqnfw}. The quantity ${\rm d}N/{\rm d}E$ denotes the differential energy spectrum of electrons and positrons produced per annihilation event. For the $e^+e^-$ annihilation channel considered in this work, we adopt the corresponding electron and positron injection spectrum from PPPC4~\cite{Cirelli:2010xx}. Since the PPPC4 spectra are tabulated for DM masses above $5~{\rm GeV}$, we extrapolate the spectrum to the sub-GeV mass range to cover the full MeV--TeV parameter space considered here.

Once produced, these particles propagate under the combined effects of diffusion and energy losses. Their steady-state distribution is determined through the transport equation
\begin{align}
\frac{\partial}{\partial t}\frac{\partial n_e}{\partial E}
=&\nabla\left[D(E,r)\nabla \frac{\partial n_e}{\partial E}\right]
+\frac{\partial}{\partial E}\left[b(E,r)\frac{\partial n_e}{\partial E}\right]
+Q(E,r),
\label{edensity_IC}
\end{align}
where $D(E,r)$ represents the spatial diffusion coefficient and $b(E,r)$ describes the total energy loss rate of electrons. We solve this equation numerically using the Green's function method implemented in the RX-DMFIT package~\cite{Ginzburg:1967zja,McDaniel:2017ppt,Colafrancesco:2005ji}.

The exact diffusion properties of charged particles in globular clusters remain uncertain. In this work we adopt a simplified form for the diffusion coefficient, parameterized as a power law in energy~\cite{Colafrancesco:2005ji,DAMPE:2022jgy,McDaniel:2017ppt}
\begin{align}
D(E)=D_0 E^{1/3},
\label{D0_IC}
\end{align}
where $D_0$ is the normalization constant. For Omega Centauri, previous studies indicate $D_0$ in the range of $D_0=10^{26}$--$10^{30}\,\rm{cm^2\,s^{-1}}$~\cite{Kar:2020coz,Wang:2023sxr}.

The total energy loss rate of the $e^\pm$ can be expressed as~\cite{McDaniel:2017ppt}
\begin{align}
b(E,r)&=b_{\rm IC\,CMB}(E,r)+b_{\rm IC\,SL}(E,r)\nonumber\\
&\quad+b_{\rm Synch}(E,r)+b_{\rm Coul}(E)+b_{\rm Brem}(E) \nonumber\\
&=(b_{\rm IC\,CMB}^0+b_{\rm IC\,SL}^0)E^2+b_{\rm Synch}^0B(r)^2E^2 \nonumber\\
&\quad+b_{\rm Coul}^0n_e\left[1+\lg\left(\frac{E/m_e}{n_e}\right)/75\right]\nonumber\\
&\quad+b_{\rm Brem}^0n_e\left[\lg\left(\frac{E/m_e}{n_e}\right)+0.36\right],
\label{energyloss_IC}
\end{align}
which includes contributions from IC scattering (from both CMB and starlight), synchrotron radiation, Coulomb interactions, and bremsstrahlung processes. The thermal electron density is taken as $n_e=0.05\,\rm cm^{-3}$~\cite{Freire:2001wq}, typical for globular cluster environments. The normalization coefficients of the energy loss rates, in units of $10^{-16},\rm GeV,s^{-1}$, are taken as~\cite{Colafrancesco:2005ji,Profumo:2010ya}
\begin{align}
b_{\rm syn}^0&\approx 0.0254,\quad 
b_{\rm IC\,CMB}^0 \approx 0.25,\nonumber\\
b_{\rm brem}^0&\approx 1.51,\quad 
b_{\rm Coul}^0 \approx 6.13.
\label{bcoefficient}
\end{align}
Also, we estimate the IC energy loss coefficient $b_{\rm IC,SL}^0(r)$ from the local starlight energy density $U_{\rm SL}(r)$. In the Thomson limit, the IC energy loss coefficient scales linearly with the target photon energy density, and thus
\begin{equation}
b_{\rm IC\,SL}^0(r)=b_{\rm IC\,CMB}^0\,\frac{U_{\rm SL}(r)}{U_{\rm CMB}},
\label{bICSL}
\end{equation}
where $U_{\rm SL}(r)$ can be calculated from the spatially resolved starlight radiation field described in Sec.~\ref{sec:starlight}. For the CMB, we adopt an energy density of $U_{\rm CMB}=0.260\,{\rm eV\,cm^{-3}}$~\cite{Hill:2018trh}.
Moreover, the magnetic field strength in Omega Centauri is estimated to be on the order of $1\sim10\,\rm{\mu G}$~\cite{Kar:2020coz}. We adopt an exponential magnetic-field profile $B(r)=B_0 e^{-r/r_{\rm c}}$, with $B_0=5\,\rm{\mu G}$ and core radius equals to the half light radius $r_{\rm c}=r_{\rm h}=7\,{\rm pc}$~\cite{Chen:2021rea,Wang:2023sxr}, to calculate the synchrotron energy loss of the $e^\pm$.

Next, the IC emissivity can be calculated through
\begin{align}
j_{\rm IC}(\nu,r)=2\int_{m_e}^{m_{\chi}}{\rm d}E
\frac{{\rm d}n_e}{{\rm d}E}(E,r)P_{\rm IC}(\nu,E,r),
\label{jIC}
\end{align}
where $P_{\rm IC}(\nu,E,r)$ is the standard inverse Compton power, describing the photon production rate at frequency $\nu$ for a single electron with energy $E$ scattering off the ambient radiation field. Detailed expressions for $P_{\rm IC}$ can be found in~\cite{McDaniel:2017ppt}.

Finally, the observable energy flux can be obtained by integrating the emissivity along the line of sight and over the solid angle of the source,
\begin{align}
\nu S(\nu)=\frac{\nu}{4\pi} \int_{\Omega}{\rm d}\Omega
\int_{\rm los} j_{\rm IC}(\nu,r)\,{\rm d}l .
\label{flux_IC}
\end{align}

\begin{figure}[htb]
\centering
\includegraphics[width=\linewidth]{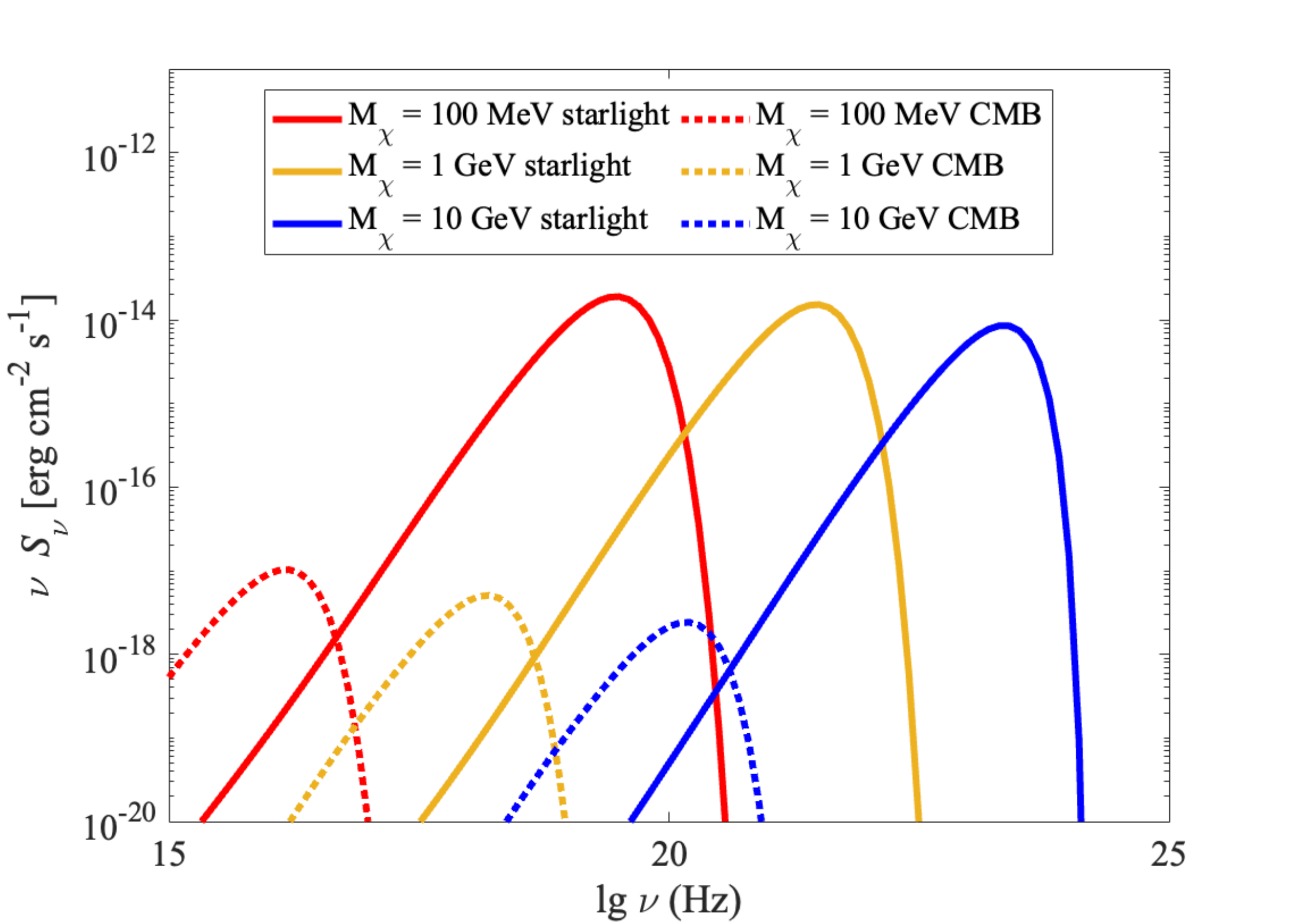}
\caption{Predicted IC flux from DM annihilation in Omega Centauri for $m_\chi = 100~\rm MeV$ (red lines), $1~\rm GeV$ (yellow lines), and $10~\rm GeV$ (blue lines). Solid lines correspond to IC flux from starlight photon, while dashed lines show the IC flux from CMB photon. Here we choose $\langle\sigma v\rangle = 10^{-30}\,\rm cm^3\,s^{-1}$ and $D_0 = 10^{28}\,\rm cm^2\,s^{-1}$. Also, we use DM distribution profile 1 in Tab.~\ref{tab:profile} as an example. Here the DM annihilation channel is $e^+e^-$.}
\label{fig:IC_SED}
\end{figure}

In Fig.~\ref{fig:IC_SED}, we present the predicted IC flux from starlight for three representative DM masses, $m_\chi = 100~\rm MeV$, $1~\rm GeV$, and $10~\rm GeV$, shown as red, yellow, and blue solid lines, respectively. Here we adopt a diffusion coefficient $D_0 = 10^{28}~\rm cm^{2}\,s^{-1}$ and an annihilation cross section $\langle\sigma v\rangle=10^{-30} ~\rm cm^{3}\,s^{-1}$. For DM distribution, we choose profile 1 in Tab.~\ref{tab:profile} as an example. The resulting IC emission spans a wide energy range, covering from X-ray to gamma-ray range. This enables us to constrain the DM annihilation cross section by combining X-ray and gamma-ray instruments, such as Chandra and Fermi-LAT. For comparison, we also present the IC emission from the CMB shown as dashed lines with the same color coding. The starlight induced IC emission exceeds the CMB induced one by about three orders of magnitude, which is expected to provide significantly stronger constraints on DM annihilation cross-section. Moreover, we note that the flux decreases with increasing DM mass, which can be caused by the stronger energy losses experienced by higher-energy electrons during their propagation.

\section{Constrains on DM Annihilation Cross-Section}
\label{sec:constraints}

In this section, we derive the upper limits on the DM annihilation cross-section $\langle\sigma v\rangle$ with observations from Chandra and Fermi-LAT. For the sub-GeV DM mass range, we utilize the flux upper limits derived from Chandra observations, where no significant extra diffuse signals have been detected. For DM masses above GeV, we perform a statistical analysis by combining the predicted DM signal with the total background model to fit the Fermi-LAT gamma-ray data.

\subsection{Constraints from Chandra}
\label{sec:chandra}

For sub-GeV DM, the IC emission from electrons scattering off starlight peaks in the X-ray band. X-ray observations therefore provide a powerful probe of this mass range.

We use Chandra observations of Omega Centauri to constrain the DM annihilation cross section $\langle\sigma v\rangle$. From the cluster core, a $2\sigma$ upper limit on the integrated luminosity of $0.5$--$4.5\,{\rm keV}$ unresolved source population is $L_X < 3 \times 10^{30}~\rm erg~s^{-1}$~\cite{Haggard:2009ts}. To simplify the analysis, we do not perform a full integration over the energy band. Instead, we require that the predicted monochromatic IC luminosity at the characteristic frequency of $5\times 10^{17}~\rm Hz$ (about $2{\rm\,keV}$) obtained from Eq.~\eqref{flux_IC} does not exceed this observed threshold. We take the core radius to be equal to the half light radius $r_{\rm c}=r_{\rm h}=7\,{\rm pc}$, corresponding to an angular radius of approximately $4.5'$. The resulting constraints, together with those derived from Fermi-LAT observations, are presented in Sec.~\ref{sec:results}, which should be regarded as conservative constraints.


\subsection{Constraints from Fermi-LAT}
\label{sec:Fermi}

In this subsection we turn to gamma-ray observations to constrain DM with masses above $\sim1$~GeV, where the IC emission shifts into the Fermi-LAT energy range. Our analysis proceeds in two steps. First, we model the astrophysical gamma-ray background of Omega Centauri. Then, we fix this background and use a $\chi^2$ profile likelihood method to set upper limits on the DM annihilation cross section 

\subsubsection{Gamma-ray Background Model}
\label{sec:Fermi_background}

The gamma-ray background from Omega Centauri consists of two distinct astrophysical components: the prompt magnetospheric radiation from millisecond pulsars (MSPs) and the IC scattering of relativistic \(e^{\pm}\) pairs injected by MSPs. Following the spectral analysis method in~\cite{Song:2021zrs}, we model the total background flux as a sum of these two components:
\begin{align}
\frac{{\rm d}N}{{\rm d}E} &= \left[ \frac{{\rm d}N}{{\rm d}E} \right]_{\rm MSP} + \left[ \frac{{\rm d}N}{{\rm d}E} \right]_{\rm IC}\nonumber\\
&= A_{\rm MSP} \left( \frac{E}{E_0} \right)^{-\Gamma_1} \exp \left( -\frac{E}{E_{\rm cut}} \right) + A_{\rm IC} \left( \frac{E}{E_0} \right)^{-\Gamma_2},
\label{eq:bg_total}
\end{align}
where $[ {\rm d}N/{\rm d}E]_{\rm MSP}$, originating from the curvature radiation of primary electrons in MSP magnetospheres, is modeled using a power law with an exponential cutoff, while \([ {\rm d}N/{\rm d}E]_{\rm IC}\) represents the high-energy tail produced by IC up-scattering of ambient starlight and CMB photons. Here \(A_{\rm MSP}\) and \(A_{\rm IC}\) are the normalizations, \(\Gamma_1\) and \(\Gamma_2\) are the spectral indices, and \(E_{\rm cut}\) is the cutoff energy.

\begin{figure}[htb]
\centering
\includegraphics[width=\linewidth]{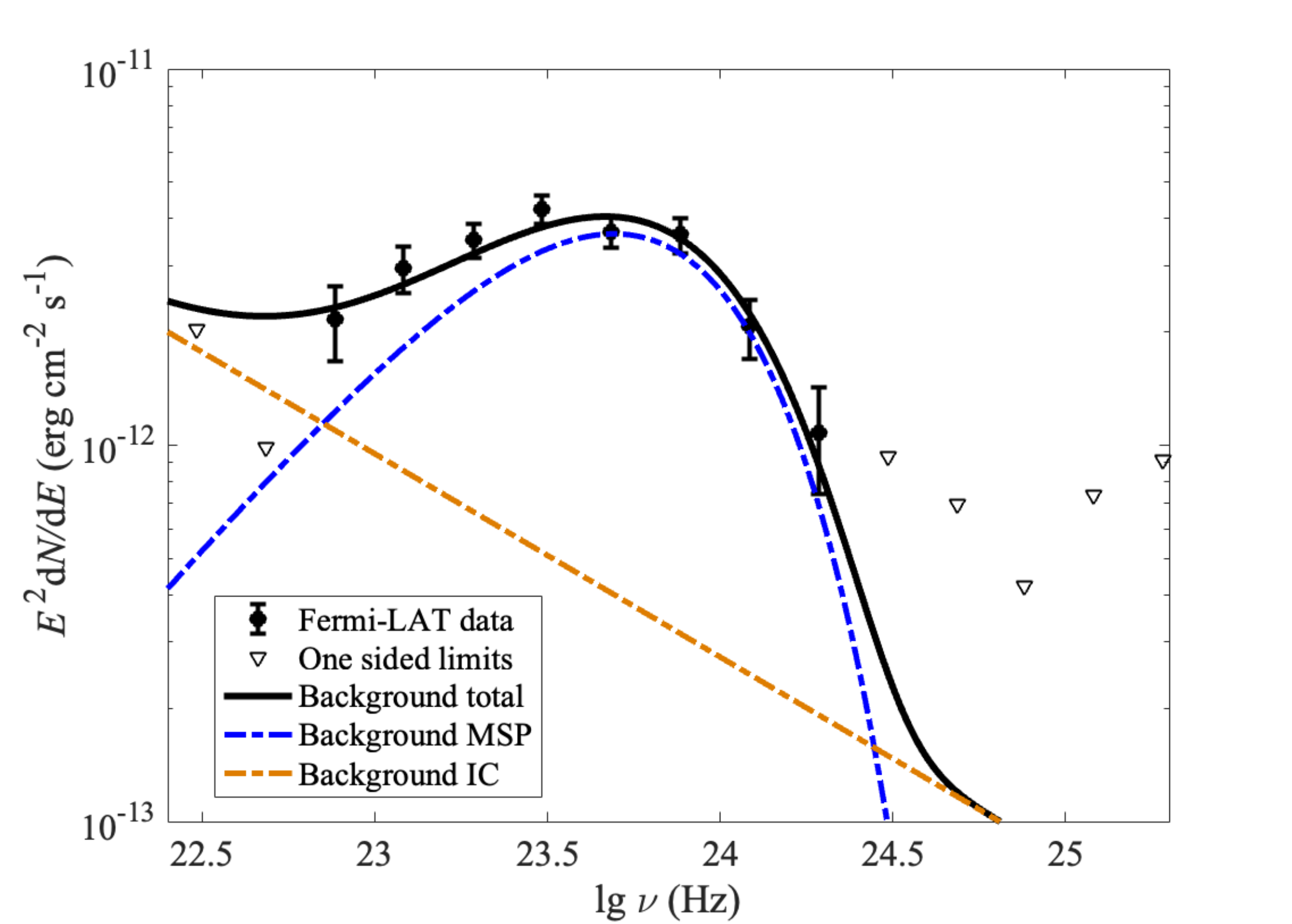}
\caption{The best-fit gamma-ray background spectrum of Omega Centauri. The black circles and open downward triangles denote the Fermi-LAT data points and upper limits, respectively. The blue dashed line shows the MSP component, and the orange dashed line shows the IC component. The total best-fit background model is shown as the black solid line.}
\label{fig:Fermi_background}
\end{figure}

To determine the background parameters, we perform a \(\chi^2\) fit to the Fermi-LAT data of Omega Centauri~\cite{Brown:2019whs}. Since the strength of the IC component is expected to be comparable to or slightly smaller than that of the MSP component~\cite{Brown:2019whs}, we impose a constraint on the ratio \(R = A_{\rm IC}/A_{\rm MSP}\) at the pivot energy \(E_0 = 1\,{\rm GeV}\), restricting it to the range \(0.3\)--\(1.0\). The best-fit gamma-ray background spectrum for Omega Centauri is shown in Fig.~\ref{fig:Fermi_background}. The black circles represent the Fermi-LAT observed data with \(1\sigma\) statistical error bars. For energy bins with low test statistic values, we show upper limits as open downward triangles. In our likelihood analysis, these upper limits are treated as one-sided constraints: the \(\chi^2\) contribution is calculated only when the theoretical model flux exceeds the observed upper limit; otherwise, the contribution is set to zero. The best-fit total background (black solid line) is composed of two primary components: the MSP magnetospheric emission (blue dashed line) and the IC emission from MSP-injected electrons (orange dashed line).

\subsubsection{Statistical Methodology for Upper Limits}
\label{sec:Fermi_limit}

To derive the constraints on the DM annihilation cross-section $\langle\sigma v\rangle$, we adopt a frequentist statistical approach based on the $\chi^2$ profile. We fix the astrophysical background to the best-fit model obtained in Sec.~\ref{sec:Fermi_background}. For a given DM mass $m_\chi$, the total predicted flux is the sum of this fixed background and the DM-induced IC component as shown in Eq.~\eqref{flux_IC}. Here we use the angular radius of approximately $4.24^\circ$, corresponding to $400\,{\rm pc}$ at a distance of $5.4\,{\rm kpc}$. The $\chi^2$ function is then constructed by comparing the model with the Fermi-LAT data. We first determine the minimum $\chi^2_{\rm min}$ by scanning the parameter space of $\langle\sigma v\rangle$. The $95\%$ C.L. upper limit on the annihilation cross-section, denoted as $\langle\sigma v\rangle_{\rm UL}$, is determined by finding the value that satisfies:
\begin{equation}
    \Delta \chi^2 = \chi^2(\langle\sigma v\rangle_{\rm UL}) - \chi^2_{\rm min} = 2.71.
\end{equation}
This procedure is repeated for each DM mass to construct the final constraint presented in Sec.~\ref{sec:results}.

\subsection{Results}
\label{sec:results}

\begin{figure}[htb]
\centering
\includegraphics[width=\linewidth]{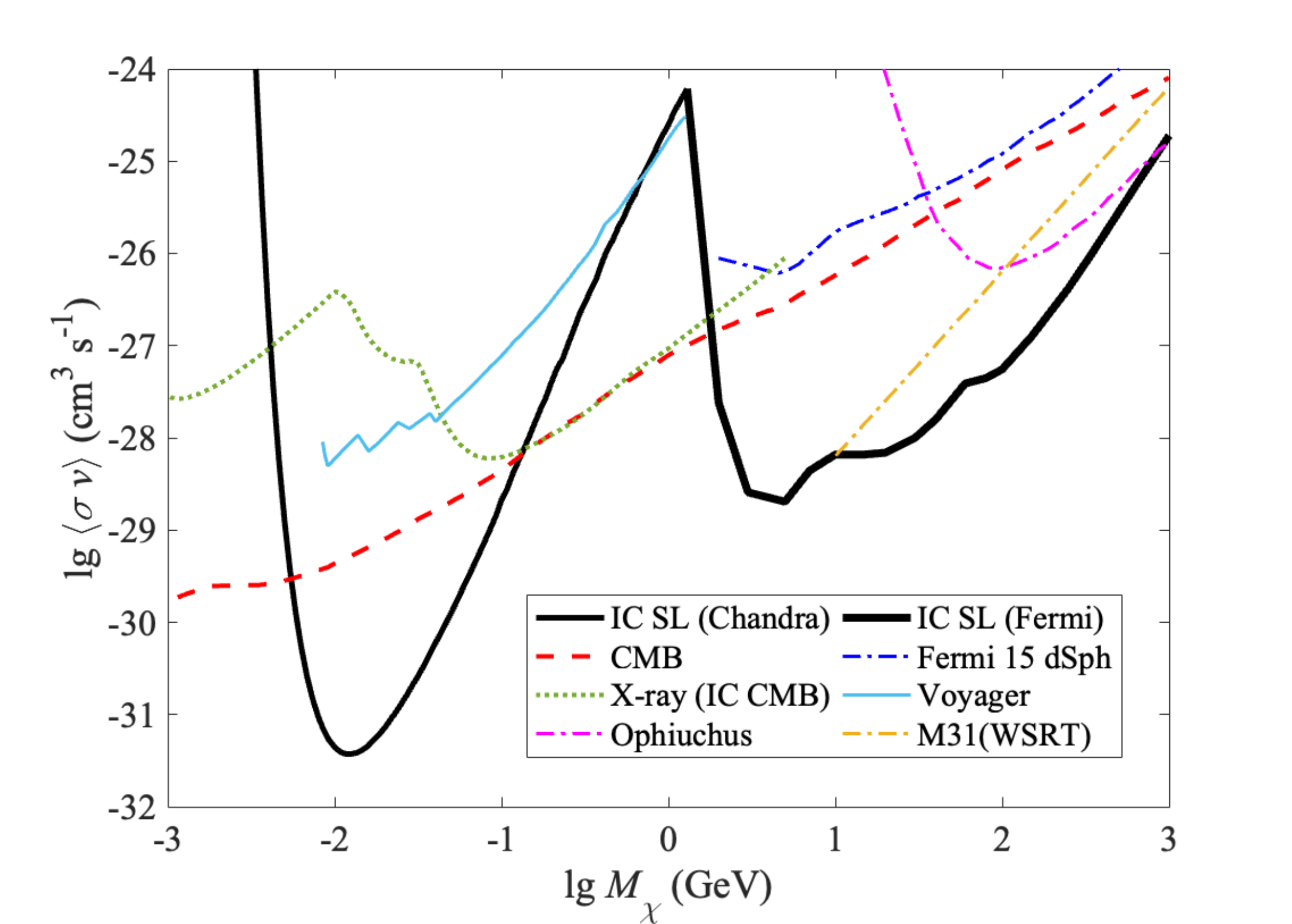}
\caption{The upper limits on $\langle\sigma v\rangle$ of the $e^+e^-$ annihilation channel} as a function of $m_\chi$. The results of this work are shown as black solid lines (thin for Chandra, thick for Fermi-LAT). Here we use Profile 1 in Tab.~\ref{tab:profile} and $D_0 = 10^{28}\,\rm{cm^2\,s^{-1}}$. For comparison, we present the limits from the CMB~\cite{Slatyer:2015jla,Leane:2018kjk} (red dashed line), Voyager~\cite{Boudaud:2016mos} (cyan solid line), X-ray from IC/CMB~\cite{Cirelli:2020bpc,Cirelli:2023tnx} (green solid line), 15 dwarf galaxies with Fermi-LAT~\cite{Baring:2015sza,Fermi-LAT:2016uux} (blue dash-dotted line), Ophiuchus~\cite{Murgia:2010sm,Yuan:2020xrz} (magenta dash-dotted line), and M31~\cite{Giessubel:2013via,Chan:2016wpu} (yellow dash-dotted line).
\label{fig:final_limits}
\end{figure}

In this subsection, we present the upper limits on the DM annihilation cross-section $\langle\sigma v\rangle$ from Chandra and Fermi-LAT observations. Here we adopt the DM density Profile 1 in Tab.~\ref{tab:profile} and a diffusion coefficient $D_0 = 10^{28}\,\rm{cm^2\,s^{-1}}$. The resulting constraints are shown as black solid lines in Fig.~\ref{fig:final_limits}, where the thin and thick lines correspond to the limits derived from Chandra and Fermi-LAT, respectively. For comparison, we also plot the limits from several other observations and channels: the CMB observation~\cite{Slatyer:2015jla,Leane:2018kjk} (red dashed line), cosmic ray measurements from Voyager~\cite{Boudaud:2016mos} (cyan solid line), X-ray from IC/CMB~\cite{Cirelli:2020bpc,Cirelli:2023tnx} (green solid line), the combined analysis of 15 dwarf galaxies by Fermi-LAT~\cite{Baring:2015sza,Fermi-LAT:2016uux} (blue dash-dotted line), and radio observations of the Ophiuchus cluster~\cite{Murgia:2010sm,Yuan:2020xrz} (magenta dash-dotted line) and M31~\cite{Giessubel:2013via,Chan:2016wpu} (yellow dash-dotted line).

As illustrated in Fig.~\ref{fig:final_limits}, our results improve upon the existing DM constraints in the MeV-to-TeV mass range. In the MeV region, the Chandra based limit provides much stronger constraints compared to the limits derived from CMB observations and X-ray observations considering only the CMB induced IC. Also, in the mass range from several GeV to the TeV scale, our Fermi-LAT constraints are also stronger than existing limits. This underscores that IC scattering off the intense starlight radiation field will serve as a powerful and indispensable tool for DM indirect searches in globular clusters, providing a unique window for DM detection.

\subsection{Impact of Parameter Uncertainties}
\label{sec:uncertainties}

The constraints derived in Sec.~\ref{sec:results} depend on the assumptions regarding the DM spatial distribution and the diffusion properties of charged particles within Omega Centauri. In this subsection, we quantify how the variation of these parameters affects the final limits on $\langle\sigma v\rangle$.

\begin{figure}
\centering 
\subfigure[]{\includegraphics[width=1\linewidth]{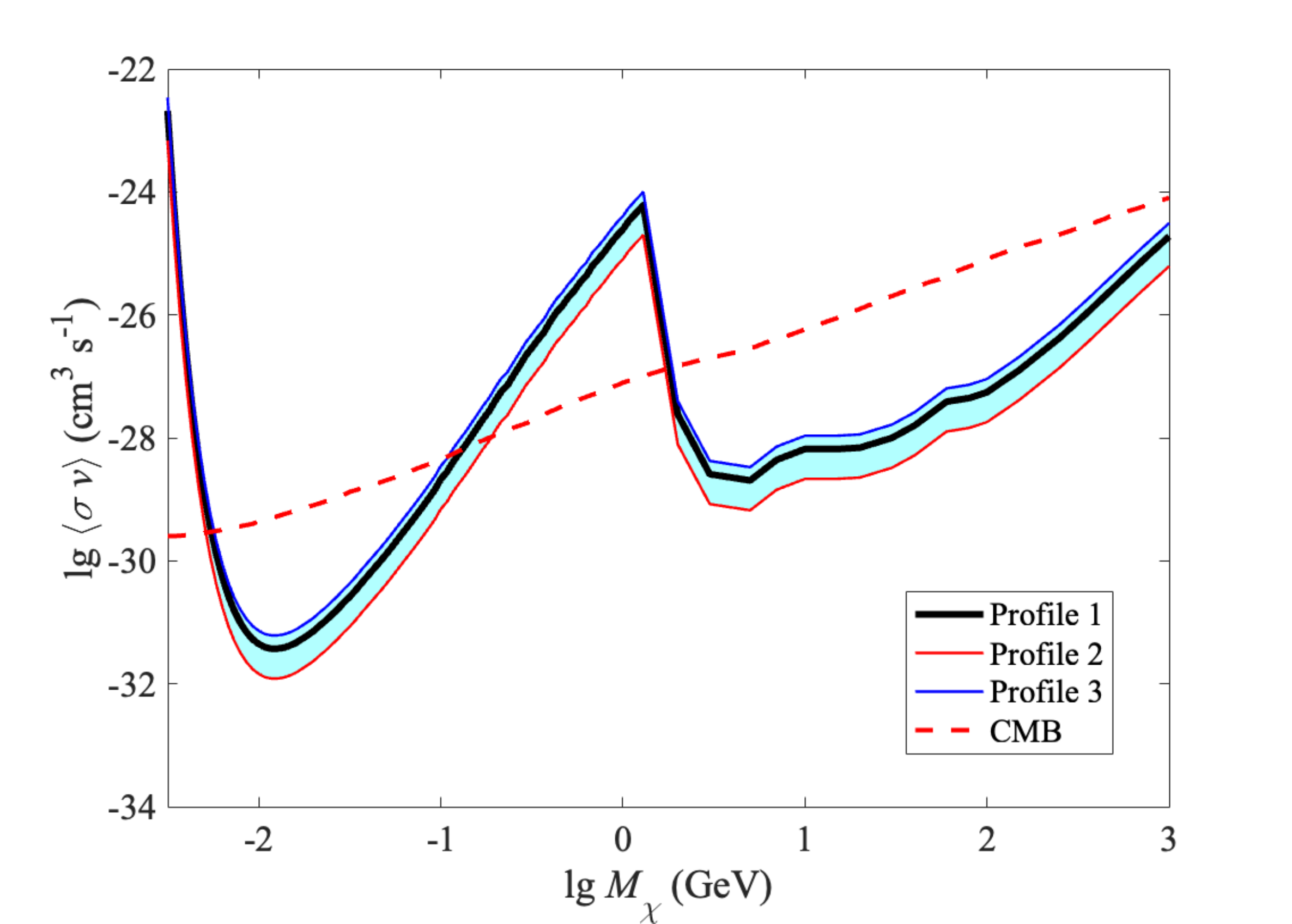}\label{fig:DMprofile}} \\
\subfigure[]{\includegraphics[width=1\linewidth]{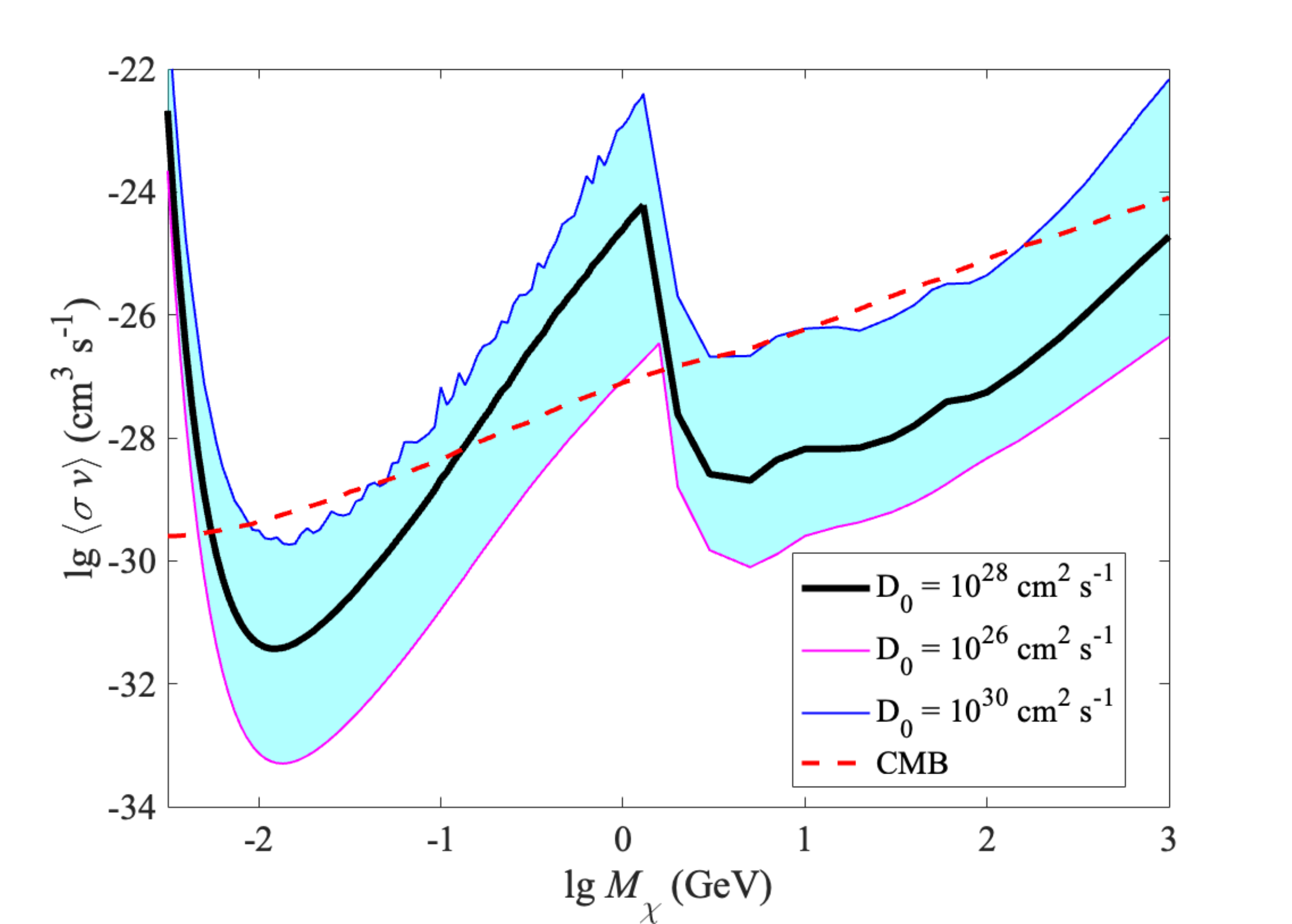}\label{fig:D0}}\\
\subfigure[]{\includegraphics[width=1\linewidth]{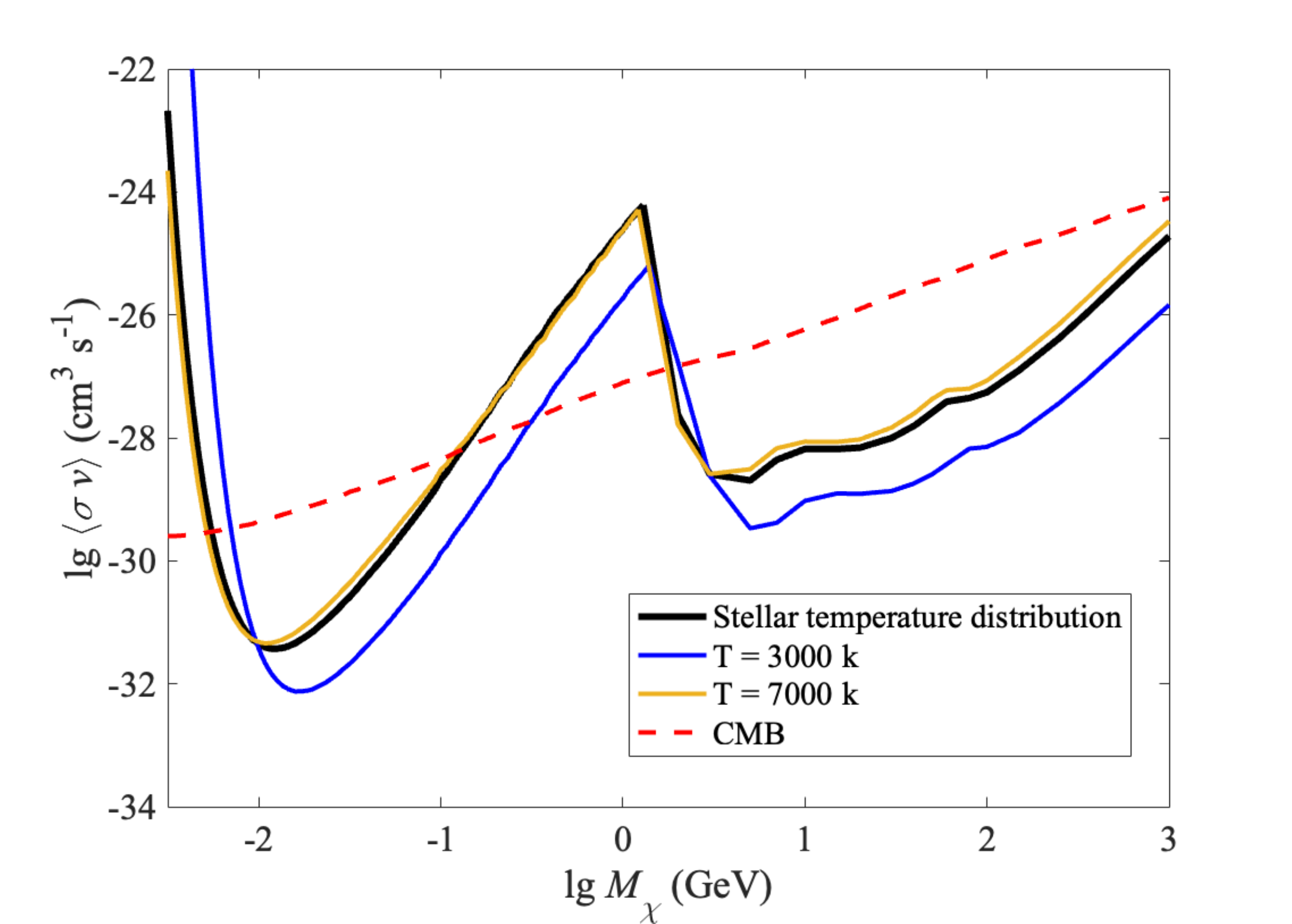}\label{fig:temperature}} 
\caption{Impact of parameter uncertainties on the constraints of the DM annihilation cross-section $\langle\sigma v\rangle$ for the $e^+e^-$ annihilation channel}. (a): Results for different DM density profile models, assuming a fixed diffusion coefficient $D_0 = 10^{28}\,\rm{cm^2\,s^{-1}}$. (b): Results for different diffusion coefficients, assuming DM Profile 1 in Tab.~\ref{tab:profile}. (c): Results obtained using the stellar temperature distribution and the two extreme cases in which all stars have effective temperatures of $3000\,\rm{K}$ or $7000\,\rm{K}$. In all three subfigures, the black lines represent our reference results shown in Fig.~\ref{fig:final_limits}.
\label{fig:uncertainty}
\end{figure}

We first consider the impact of the DM density distribution. As illustrated in Tab.~\ref{tab:profile}, we adopt three representative NFW profiles. The results are shown in Fig.~\ref{fig:DMprofile}. We find that the upper limits on the annihilation cross section approximately scale inversely with the $J$-factor. The $J$-factors considered in our work range from approximately $10^{23}\,{\rm GeV^2\,cm^{-5}}$ (Profile 3) to $5\times10^{23}\,{\rm GeV^2\,cm^{-5}}$ (Profile 2), while previous studies have reported values ranging from approximately $10^{22}$ to $10^{24}\,{\rm GeV^2\,cm^{-5}}$~\cite{Evans:2021bsh}, the uncertainty in the actual DM content could lead to larger variations in final results.

Also, we perform an uncertainty analysis of diffusion coefficient $D_0$ in Fig.~\ref{fig:D0} with three representative values: $D_0 = 10^{26}$ (red lines), $10^{28}$ (black lines), and $10^{30}\,\rm{cm^2\,s^{-1}}$ (yellow lines). We find that the constraints fluctuate by approximately two orders of magnitude in each direction relative to our benchmark case ($10^{28}\,\rm{cm^2\,s^{-1}}$). Specifically, the smallest diffusion coefficient ($10^{26}\,\rm{cm^2\,s^{-1}}$) produces the strongest results, while the largest ($10^{30}\,\rm{cm^2\,s^{-1}}$) results in the weakest. This behavior arises because a smaller $D_0$ confines the injected $e^+e^-$ pairs in the center region of the cluster. Thus, a higher fraction of the electron energy is converted into IC photons before they escape the diffusion regions. However, even under the most conservative assumptions with $D_0 = 10^{30}\,\rm{cm^2\,s^{-1}}$, the constraints remain the same level as those derived from the CMB in several tens MeV and several GeV to hundreds of GeV.

We further investigate the uncertainty of the stellar radiation field, as shown in Fig.~\ref{fig:temperature}. In addition to our fiducial stellar temperature distribution, we consider two extreme cases in which all stars have the same effective temperature of either \(3000\,\rm{K}\) (blue lines) or \(7000\,\rm{K}\) (yellow lines). Different starlight distributions lead to slight horizontal shifts in the strongest position of the constraint curve. A higher overall stellar temperature shifts the strongest position slightly toward lower DM masses, whereas a lower temperature shifts it toward higher masses. The overall differences in constraint strength remain within one order of magnitude. The \(3000\,\rm{K}\) case yields particularly strong constraints. However, as we discussed in Sec.~\ref{sec:starlight}, the \(3000\,\rm{K}\) situation should be regarded only as an extreme illustrative case, rather than a physically reliable constraint.
In summary, although the constraints are subject to uncertainties in the astrophysical parameters, our results remain relatively strong across the parameter choices considered. These results confirm the importance of globular clusters as DM probes.

\section{Conclusion}
\label{sec:conclusion}

In this work, we use multi-wavelength observations from Chandra and Fermi-LAT to constrain the DM annihilation cross section of the $e^+e^-$ annihilation channel.
We adopt a detailed spatially resolved starlight radiation field derived from stellar population synthesis and the observed luminosity profile of the globular cluster. 
Our calculation shows that the starlight in the cluster core provides a much more efficient target for IC scattering from DM-injected \(e^+e^-\) pairs, leading to an IC flux more than two orders of magnitude higher than that from CMB. By combining the X-ray flux upper limits from Chandra with the gamma-ray data from Fermi-LAT, we derive new constraints on \(\langle\sigma v\rangle\).
Our results demonstrate that Omega Centauri provides stringent limits in the MeV to TeV mass range. Under typical propagation parameters, the limits improve upon existing results by nearly two orders of magnitude in the tens-of-MeV region, exceeding \(\langle\sigma v\rangle \sim 10^{-31}~{\rm cm^3\,s^{-1}}\). Also, our constraints from Fermi-LAT at several GeV to TeV exceed existing results. The derived constraints depend on several uncertain factors, including the diffusion coefficient and the DM density profile. In particular, the DM content of Omega Centauri remains debated, introducing additional uncertainty into our results. Nevertheless, under conservative assumptions, our limits in both the tens-of-MeV and several-GeV-to-TeV regions remain comparable to those derived from CMB observations.
Our results underscore that the dense stellar environments of globular clusters, especially those with possible galactic remnant origins, such as Omega Centauri, are excellent targets for indirect DM searches. 

\vspace{.5cm}
\noindent {\bf Acknowledgements}
This work is supported by the National Key R\&D Program of China (Grant No. 2022YFF0503304), the National Natural Science Foundation of China (Grant Nos. 12373002, 12220101003, and 11773075),  the Youth Innovation Promotion Association of Chinese Academy of Sciences (Grant No. 2016288) and Project of Aba Teachers  College (Grant No. AS-RCZX2025-08).

{\small
\bibliographystyle{spphys}
\balance
\bibliography{ref}
}

\end{document}